\documentclass[twocolumn,showpacs,preprintnumbers,amsmath,amssymb]{revtex4}
\usepackage{graphicx}% Include figure files
\usepackage{dcolumn}% Align table columns on decimal point
\usepackage{bm}% bold math

\begin{document}

\preprint{}

\title{Gallium vacancy and the residual acceptor in undoped GaSb studied by positron lifetime spectroscopy and photoluminescence}% Force line breaks with \\

\author{C. C. Ling}
 \email{ccling@hku.hk}

\author{W. K. Mui}
\author{C. H. Lam}
\author{C. D. Beling}
\author{S. Fung}
\author{M. K. Lui}
\affiliation{Department of Physics, The University of Hong Kong, Pokfulam Road, Hong Kong, China}

\author{K. W. Cheah}
\author{K. F. Li}
\affiliation{Department of Physics, Hong Kong Baptist University, Kowloon Tong, Hong Kong, China}
\author{Y. W. Zhao}
\affiliation{Material Science Centre, Institute of Semiconductors, Chinese Academy of Sciences, Beijing, China}
\author{M. Gong}
\affiliation{Department of Physics, Sichuan University, Chengdu, China}

\date{\today}% It is always \today, today,

\begin{abstract}
Positron lifetime,  Photoluminescence and Hall measurements  were performed to study undoped p-type gallium antimonide materials.  A 314ps lifetime component, attributed to $V_{Ga}$ related defect, was identified in the positron lifetime measurement.  In the PL measurement, a $778meV$ and a $797meV$ peaks were observed.  Isochronal annealing studies were performed and at the temperature of $300^{o}C$, both the 314ps positron lifetime component and the two PL signals disappeared, which gives a clear and strong evidence for their correlation.  However, the hole concentration ($\sim 2\times 10^{17}cm^{-3}$) was observed to be constant throughout the whole annealing temperature range up to $500^{o}C$.  Contradictory to general belief, this implies, at least for samples with annealing temperatures above $300^{o}C$, the Ga vacancy is not the acceptor responsible for the p-type conduction.
\end{abstract}

\pacs{61.72Ji,78.70Bj,78.55.Cr}% PACS, the Physics and Astronomy
\maketitle

Gallium Antimonide is the basic material for a variety of lattice parameter matched III-V compounds having band gaps ranging from 0.3eV to 1.58eV (corresponding to wavelengths of 0.8 to 4.3$\mu$m) (See reviews \cite{MilnesRev,DuttaRev}).  Thus, GaSb and its lattice matched materials are capable of fabricating optoelectronic devices operating in a wide range of wavelength, high frequency devices and thermophotovoltaic devices.  Undoped GaSb materials are p-type conducting having a hole concentration of $10^{16}-10^{17}cm^{3}$.  For the PL studies of such material \cite{Allegre,Jakowetz,Lee,Dutta}, a luminescence signal called band A (peaking at 778meV) is commonly found irrespective of growth method.  This signal and also the residual acceptor were related to Ga in excess and were generally considered being due to the $V_{Ga}Ga_{Sb}$ defect \cite{MilnesRev,DuttaRev,Allegre,Jakowetz,Lee,Dutta,Meulen}, though no further direct evidence for this assignment had been observed.  In this study, we have studied undoped p-type  GaSb materials using positron lifetime spectroscopy, photoluminescence and Hall measurement, particular with an intension to study the correlation between the PL signals, the Ga vacancy and the hole concentration.

Samples were cut from two different LEC grown undoped GaSb wafers (namely called GaSb042Un with  $p=2.5\times 10^{17}cm^{-3}$ and GaSb342Un with $p=2.0\times 10^{17}cm^{-3}$) commercially purchased from the MCP Wafer Technology Ltd.  Each of the isochronal annealing process was carried out in forming gas ($N_{2}:H_{2}=80\% : 20\%$) for a period of 30 minutes.  After the annealing, the samples were moved to a room temperature region while still kept in the forming gas atmosphere before they were cooled down and removed from the furnace.  Details of positron lifetime measurement were as reported in \cite{LingGaSb}.  The 4-million-count positron lifetime spectra collected with a fast-fast positron lifetime spectrometer having a resolution of fwhm=230ps were analyzed by the source code POSITRONFIT \cite{positronfit}, in which the measured spectra were fitted by the equation: $\sum_{i}I_{i}exp(-\lambda_{i} t)$,  where $I_{i}$ and $\tau_{i} =1/\lambda_{i}$ are the intensity and the characteristic positron lifetime of the corresponding annihilation site, with the consideration of the instrumental convolution and background contribution.  The PL measurements were performed at 10K and the details can be found in Mui et al \cite{Mui}.  

Before discussing the positron lifetime results of the undoped GaSb samples, it is interest to note our previous results of a positron lifetime study on a heavily Zn-doped GaSb sample (GaSb098Zn) having $p=3.28\times 10^{18}cm^{-3}$ reported in Ref.\cite{LingGaSb}.  Referring to Ref.\cite{LingGaSb}, a two-component model was found to well describe the spectra in the as-grown sample.  The long lifetime component having characteristic lifetime of $\tau_{2}=318\pm 7ps$ was attributed to positrons annihilating at $V_{Ga}$ related defects.  Isochronal annealing studies of the Zn-doped sample indicated there were two annealing stages, namely starting at $300^{o}C$ and $580^{o}C$, and the average positron lifetime is also plotted in figure 1 for reference.  According to Ref.\cite{LingGaSb}, the drop in average lifetime at $300^{o}C$ in figure 1 was shown to correspond with an increase of the long lifetime component from 318ps to 379ps and a decrease of the long lifetime intensity from 50\% to 15\%.  This was argued to be related to the annealing out of the $V_{Ga}$ related defect and the formation of a new defect having a lifetime of 379ps.  At the $580^{o}C$ annealing stage, the lifetime spectrum became containing only one component with value of 267ps.  In Ref.\cite{LingGaSb}, it was also shown that the GaSb bulk lifetime obtained from considering the temperature varying experiment of the as-grown sample and the isochronal annealing studies agreed with each other at a value of 267ps.

\begin{figure}
\centerline{\scalebox{1.2}{\includegraphics{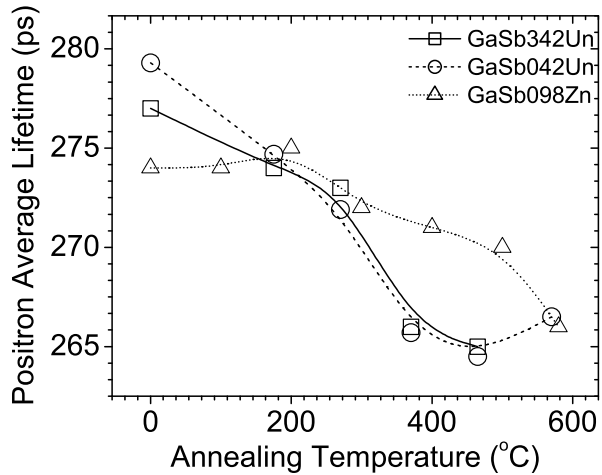}}}
\caption{Average positron lifetime of the undoped GaSb samples GaSb042Un (circle) and GaSb342Un (square) as a function of the annealing temperature.  The average lifetime data of the Zn-doped sample GaSb098Zn (triangle) presented in Ling et al is also included here for comparison.}
\end{figure} 

\begin{figure}
\centerline{\scalebox{1.1}{\includegraphics{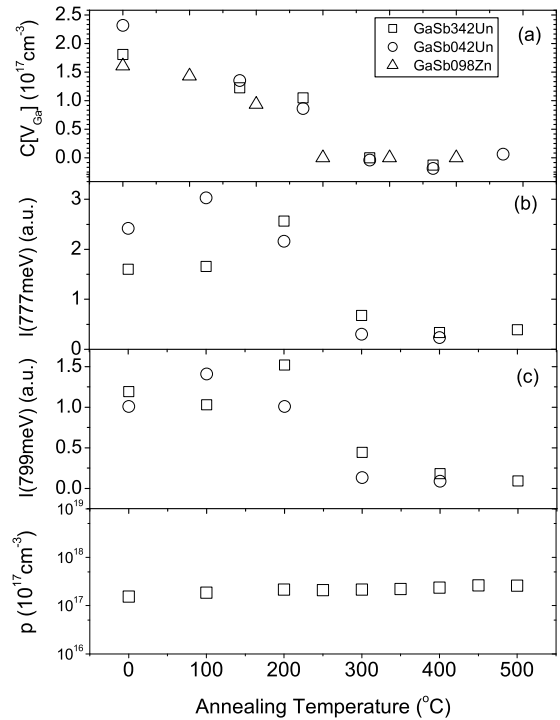}}}
\caption{(a)  Concentration of the $V_{Ga}$ defects in the undoped GaSb samples calculated from the positron lifetime data as a function of the annealing temperature;  (b) and (c)  Peak intensities and positions of the two significant PL signals observed in the undoped GaSb samples as a function of the annealing temperature; (d)  Hole concentration as a function of the annealing temperature.}
\end{figure}

The positron average lifetime as a function of the annealing temperature measured at room temperature for the two undoped GaSb samples are shown in figure 1 and they have very similar behaviors.  There is an annealing stage at temperature of $300^{o}-400^{o}C$, for which the average lifetime drops to abut 266ps.  The spectra of the undoped samples were also fitted with the source code POSITRONFIT.  It was found that at an annealing temperature lower than $370^{o}C$, a two component model was found to offer a good fit to the spectra and the long lifetime component was constant at 314ps.  At annealing temperatures higher than $370^{o}C$, a single component model with lifetime of 266ps was employed to give good fit to the data.  This implies the positron trapping centre having a lifetime of 314ps was annealed out at about $370^{o}C$.  Above this annealing temperature, no positron trapping process was observed and the bulk lifetime is 266ps, which is coincident with that in Ref.\cite{LingGaSb}.  The 314ps lifetime component identified in the undoped samples is also close to the 318ps found in the Zn-doped sample, which was attributed to Ga vacancy related defect \cite{LingGaSb}.  Furthermore, the annealing behaviors of the 314ps component in the undoped samples and the 318ps component in the Zn-doped sample are very similar (annealing behavior of $V_{Ga}$ related defect in Zn-doped sample published in Ref.\cite{LingGaSb} is also included as symbol triangle in figure 2(a) for reference).  It is thus plausible to conclude the 314ps component found in the undoped GaSb samples at annealing temperatures lower than $370^{o}C$ is due to $V_{Ga}$ related defect and it disappears at an annealing temperature of $370^{o}C$.

With the simple trapping model [e.g. Ref.\cite{KrauseBook}], the positron trapping rate of the $V_{Ga}$ defect $\kappa$ is related to the characteristic defect lifetime $\tau_{d}$, the bulk lifetime $\tau_{bulk}$ and the average lifetime $\tau_{ave}$ as: $\kappa=[\tau_{ave}-\tau_{b}]/[(\tau_{b}(\tau_{d}-\tau_{ave})]=\mu \times c$, where $c$ and $\mu$ are the concentration and the specific positron trapping coefficient of $V_{Ga}$.  Although the precise value of $\mu$ for the Ga vacancy in GaSb is still not available, based on  the specific positron coefficient values of $V_{Ga}^{-}$ in GaAlSb and $V_{Ga}^{-}$ in GaAs,  the specific positron trapping coefficient for $V_{Ga}$ in GaSb was estimated to be $\mu \sim 2\times 10^{14}s^{-1}$ \cite{LingGaSb}.  The calculated  Ga vacancy concentrations of the two undoped GaSb samples as a function of the annealing temperature are shown in figure 2(a).  It is clearly seen that the Ga vacancy in all the three samples has a concentration of $\sim 10^{17}cm^{-3}$ and disappears at the temperature range of $300^{o}-400^{o}C$.

The PL spectra of the two undoped sample are very similar and those of GaSb042Un annealed at different temperatures are shown in figure 3.  From the figure, two dominant luminescence peaks were observed to be at about 780meV and  800meV.  At the low energy shoulder of the 780meV peak, there is another weak luminescence signal at about 760meV.  It is also obvious that the PL signals were greatly reduced by the $300^{o}C$ annealing.  The PL spectra were fitted with the superposition of three Gaussians and all the fitted curves have excellent values of chi-square.  The fitted peak positions of the two dominant signals were found to be constants at different annealing temperatures with values of  $777.4\pm 0.7meV$ and $798.8\pm 1.7meV$ for GaSb042Un, and $778.3\pm 1.5eV$ and $794.7\pm 2.6eV$ for GaSb342Un.   The intensities of the two peaks for both undoped samples as a function of the annealing temperature are shown in figure 2(b) and (c).  One notes that the annealing behavior of the Ga vacancy related defect detected by the positron lifetime technique is effectively the same as that of the two PL signals.  This is a clear and strong evidence suggesting that the 314ps positron lifetime component, the 778meV and the 797meV PL signals are originated from the same $V_{Ga}$ related defect.   

\begin{figure}
\centerline{\scalebox{1.3}{\includegraphics{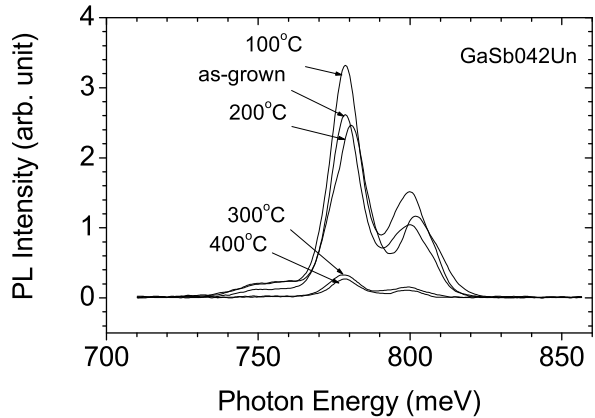}}}
\caption{Photoluminescence spectra of the undoped sample GaSb042Un annealed at different at different temperatures.}
\end{figure}

The 778meV PL signal is commonly observed in most of the p-type GaSb materials and is known as the band A.  The residual acceptor responsible for p-type conduction of undoped material and the band A PL signal  were observed to be related to the Ga excess \cite{Allegre,Jakowetz,Lee,Meulen}.  The residual acceptor is usually attributed to the $V_{Ga}Ga_{Sb}$ defect \cite{Meulen} and the band A PL signal is usually associated to CB or donor to $V_{Ga}Ga_{Sb}$ transition \cite{Allegre,Jakowetz,Lee,Dutta}.  The other dominant PL peak lines 795 to 799meV seen in the present study has been previously reported and attributed to exciton bound to neutral $V_{Ga}Ga_{Sb}$ \cite{Jakowetz,Lee,Dutta}.  The present observed evidence for the correlation between the two dominant PL peaks (778meV and 797meV) and the Ga vacancy as seen from the positron lifetime signal strongly supports the general believed physical processes for luminescence.

In order to study the correlation between the Ga vacancy and the hole concentration, the hole concentrations of the GaSb342Un sample annealed at different temperatures were measured by Hall measurement at room temperature (results shown in figure 2(d)).  The hole concentration is about $2\times 10^{17}cm^{-3}$ at all different annealing temperatures, which is consistent with the observation of stable hole concentration upon annealing reported by Effer and Etter \cite{Effer}.  In the work of Van Der Meueln \cite{Meulen}, it was pointed out the residual acceptor was related to the Ga excess and contained a vacancy in its structure.  It was further argued that because of its lack of mobility and stability upon annealing, the residual acceptor must be a $V_{Ga}Ga_{Sb}$.  However, from the annealing behavior of the lifetime signal and the hole concentration shown in figure 2(a) and (d), at least for undoped samples annealed at $300^{o}C$ ($V_{Ga}$ anneals out) or above, the p-type conduction is not originated from the Ga vacancy, but is from another acceptor.  This observation somehow contradicts the general belief that $V_{Ga}Ga_{Sb}$ is always the acceptor contributing the holes to the valance band for undoped GaSb.

\acknowledgments
This study was financially supported by the Research Grant Council, Hong Kong Special Administrative Region, China (Project No. HKU7134/99P) and the CRCG, The University of Hong Kong. 

%*****************************Reference*********************************** 


\begin{references} 
\bibitem{MilnesRev} A.G. Milnes and A.Y. Polyakov, Solid State Electronics \underline{36}, 803 (1993). 

\bibitem{DuttaRev} P.S. Dutta and H.L. Bhat, J. Appl. Phys. \underline{81}, 5821 (1997).

\bibitem{Allegre} J. All\'{e}gre and M. Av\'{e}rous, Inst. Phys. Conf. Ser. No. 46: Chapter 5, 379 (1979).

\bibitem{Jakowetz} W. Jakowetz, W. R\"{u}hle, K. Breuninger and M. Pilkuhn, Phys. Stat. Solidi A \underline{12}, 169 (1972).

\bibitem{Lee} M. Lee, D.J. Nicholas, K.E. Singer and B. Hamilton, J. Appl. Phys. \underline{59}, 2895 
(1986).

\bibitem{Dutta} P.S. Dutta, K.S.R. Koteswara Rao, H.L. Bhat and V. Kumar, Appl. Phys. A \underline{61}, 149 (1995).

\bibitem{Meulen} Y.J. Van Der Meulen, J. Phys. Chem. Solids \underline{28}, 25 (1967).

\bibitem{LingGaSb} C. C. Ling, S. Fung and C. D. Beling, Phys. Rev. B \underline{64}, 75201 (2001).

\bibitem{Mui} W. K. Mui, M. K. Lui, C. C. Ling, C. D. Beling, S. Fung, K. W. Cheah, K. F. Li and Y. W. Zhao, presented in Material Science Society Fall Meeting 2001 and proceeding to be published.

\bibitem{positronfit} 
P. Kirkegaard, N.J. Pederson and M. Eldrup, PATFIT-88; P. Kirkegaard,
M. Eldrup, O.E. Mogenson, N.J. Pederson, Comput. Phys. Commun.
\underline{23}, 307 (1981).

\bibitem {KrauseBook} R. Krause-Rehberg and H.S. Leipner, "Positron Annihilation in Semiconductors, 
Defect Studies", Springer Series in Solid-State Sciences 127, Springer-Verlag, (1999), page 85.

\bibitem{Effer} D. Effer and P. J. Etter, J. Phys. Chem. Solids \underline{25}, 451 (1964).

\end{references}
\end{document}